# The Accidental Detection Index as a Fault Ordering Heuristic for Full-Scan Circuits

Irith Pomeranz[1]
School of Electrical & Computer Eng.
Purdue University
W. Lafayette, IN 47907, U.S.A.

and

Sudhakar M. Reddy[2]
Electrical & Computer Eng. Dept.
University of Iowa
Iowa City, IA 52242, U.S.A.

**Abstract**

*We investigate a new fault ordering heuristic for test generation in full-scan circuits. The heuristic is referred to as the accidental detection index. It associates a value $ADI(f)$ with every circuit fault $f$. The heuristic estimates the number of faults that will be detected by a test generated for $f$. Fault ordering is done such that a fault with a higher accidental detection index appears earlier in the ordered fault set and targeted earlier during test generation. This order is effective for generating compact test sets, and for obtaining a test set with a steep fault coverage curve. Such a test set has several applications. We present experimental results to demonstrate the effectiveness of the heuristic.*

## 1. Introduction

Dynamic test compaction procedures for full-scan circuits [1]-[4] typically attempt to generate test vectors that detect as many faults as possible. In this way they reduce the number of test vectors required for detecting all the detectable circuit faults. Dynamic compaction heuristics added to a test generation procedure typically increase the test generation time significantly, to several times the original run time of the test generation procedure without the dynamic compaction heuristics.

Some of the dynamic test compaction heuristics can be expected to increase the run time. For example, using the unspecified values of a test vector to detect secondary target faults [1] implies that in the worst case, for a circuit with $n$ faults, there will be $O(n)$ test generation attempts for every vector instead of a single attempt when secondary target faults are not used. Other dynamic compaction heuristics do not increase the worst-case complexity of test generation. One such heuristic is fault ordering. Under a fault ordering heuristic, the faults are placed in the set of target faults $F$ in a certain order, and the test generation procedure considers the faults in the order they appear in $F$. For example, in [2], the fault ordering heuristic uses maximal sets of independent faults in fanout free regions [5]. Faults in larger sets are placed at the beginning of $F$ to ensure that the first tests generated are necessary for detecting all the faults.

In this work we investigate a new fault ordering heuristic. We refer to the heuristic as the *accidental detection index*. An accidental detection index is associated with every circuit fault $f$, and it estimates the number of faults that will be detected by a test generated for $f$. This has several applications.

(1) For dynamic test compaction, a fault with a higher accidental detection index should be targeted earlier in order to ensure that each new test vector added to the test set detects as many faults as possible. Similar to other dynamic compaction heuristics, this reduces the final test set size.

(2) The procedures of [6] and [7] reorder a test set so as to obtain a fault (or defect) coverage curve which is as steep as possible. With every additional test, a steeper fault coverage curve has a higher fault coverage. The motivation for studying this problem is as follows. The reordered test set is useful if the test set is too large to fit in the tester memory and it is necessary to remove some tests in order to avoid multiple loads of the tester memory. Removing the last tests of a reordered test set with a steeper fault coverage curve reduces the fault coverage by a smaller amount. Similarly, removing the last tests of a reordered test set can be used to reduce the test application time while minimally reducing the fault coverage. In addition, with a reordered test set, more defects are expected to be detected by the earlier tests. Thus, an appropriate reordering of the test set reduces the time a defective chip is expected to spend on a tester until the defect is detected. The method of [6] reorders the test set based on a-posteriori probabilities for tests in the test set to detect defective chips. The method of [7] uses the results of $n$-detection fault simulation, and does not require any a-posteriori information. Using the method of [7], tests that detect larger numbers of faults appear earlier in the reordered test set, satisfying the requirement for a steep fault coverage curve. If test generation is performed such that faults with a higher accidental detection index are targeted earlier, a steeper fault coverage curve is expected even without reordering the test set. In addition, the test vectors obtained in this way are expected to be more effective in

1. Research supported in part by SRC Grant No. 2004-TJ-1244.
2. Research supported in part by SRC Grant No. 2004-TJ-1243.

obtaining a steeper fault coverage curve than test vectors obtained without the accidental detection index heuristic.

In this work we compute the accidental detection index based on a given set of input vectors $U$ (a small set of random vectors that detects a certain percentage of the circuit faults). We order the set of faults *including the faults detected by* $U$. In other words, we do not drop the faults detected by $U$ from the set of target faults $F$. This is done for the following reasons. First, most of the test generation effort is expended for the detection of hard-to-detect faults, which are not detected by $U$. Thus, dropping the faults detected by $U$ from $F$ is not likely to reduce the test generation time. Second, a test for a fault with a high accidental detection index that is detected by $U$ may accidentally detect a hard-to-detect fault and save the test generation effort otherwise required to detect it. More important, random input vectors are not desirable if the goal is to generate a compact test set or achieve a steep fault coverage curve. To achieve these goals it is better to perform test generation considering all the target faults.

Fault ordering was considered earlier in [8]-[10]. In [8] and [9], fault ordering is done to speed up fault simulation of synchronous sequential circuits by grouping together faults that cause similar events. Here, we are not concerned with the events that a fault causes during fault simulation. In [10], fault ordering is done to support the generation of compact test sequences for synchronous sequential circuits by targeting hard-to-detect faults first. The accidental detection index defined here addresses more directly the issue of identifying faults whose tests will detect large numbers of other faults. These may or may not be hard-to-detect faults. Moreover, for the application of steepening the fault coverage curve, we prefer to place the hard-to-detect faults (the faults not detected by $U$) at the end of the ordered set of faults since their accidental detection index is unknown and may be low.

The paper is organized as follows. In Section 2 we define the accidental detection index. In Section 3 we discuss fault ordering based on the accidental detection index. In Section 4 we present the results of test generation using the fault orders of Section 3. We consider the two applications of dynamic test compaction and generation of test sets with steep fault coverage curves.

## 2. Accidental detection index

In this section we define the accidental detection index $ADI(f)$ for a fault $f$ and provide a method for estimating its value.

The accidental detection index of a fault $f$ is defined so as to capture the number of faults that will be detected by a test vector generated for $f$. Except for $f$, all the other faults detected by the test vector are detected accidentally. Since $f$ may have several test vectors, a conservative definition of the accidental detection index of a fault $f$ considers the minimum number of faults that will be detected by a test vector generated for $f$. It is also possible to use the average value, where the average is computed over all the test vectors that detect $f$. We take the conservative approach and define the accidental detection index of a fault $f$ as the minimum number of faults that will be detected by a test generated for $f$.

Computation of the exact value of the accidental detection index of a fault $f$ will have a high computational complexity. In this work, we compute an estimate for the accidental detection index based on a set of input vectors $U$. In our experiments, $U$ is a set of random vectors of limited size $N$. We discuss the selection of $N$ later.

We denote by $F$ the set of target faults (single stuck-at faults in our case). We denote by $F_U$ the subset of $F$ detected by $U$. We compute an accidental detection index $ADI(f)$ only for a fault $f \in F_U$, i.e., only for faults detected by $U$. For a fault $f \in F-F_U$ that is not detected by $U$ the accidental detection index is by definition zero, i.e., $ADI(f) = 0$ for $f \in F-F_U$. For $f \in F_U$, we include $f$ itself in the count of faults for the accidental detection index. Therefore, $ADI(f) \geq 1$ for every $f \in F_U$.

In order to compute $ADI(f)$ for $f \in F_U$, we simulate the faults in $F_U$ under $U$ without fault dropping and find for every fault $f \in F_U$ the subset of input vectors $D(f) \subseteq U$ that detect $f$. In addition, we find for every input vector $u \in U$ the number of faults detected by $u$. This number is denoted by $n_{det}(u)$. Instead of fault simulation without fault dropping it is also possible to use $n$-detection fault simulation to estimate $n_{det}(u)$ for every $u$.

Let $D(f) = \{u_1, u_2, \cdots, u_m\}$. For every $u_i \in D(f)$ we have the number $n_{det}(u_i)$ of faults that will be detected if $u_i$ is generated for detecting $f$. The minimum value of $n_{det}(u_i)$ over all $u_i \in D(f)$ is an estimate of the minimum number of faults that will be detected accidentally by a test generated for $f$. We use this value as the accidental detection index of $f$, i.e., we set

$ADI(f) = \min\{n_{det}(u_i) : u_i \in D(f)\}$ for $f \in F_U$.

The following example demonstrates this definition. We consider the combinational logic of MCNC finite-state machine benchmark *lion*. The circuit has four inputs and 40 single stuck-at faults included in the set of target faults $F$. We include all the 16 input vectors of the circuit in the set $U$. All the faults in $F$ are detected by $U$ and included in $F_U$. The input vectors in $U$ detect the numbers of faults shown in Table 1. For every $u \in U$ we show in Table 1 the value of $n_{det}(u)$. The vector $u$ is given by its decimal representation.

Next, we consider several of the faults of *lion* and compute for them the accidental detection index $ADI(f)$.

**Table 1: Input vectors of** *lion*

| $u$ | 0 | 1 | 2 | 3 | 4 | 5 | 6 | 7 |
|---|---|---|---|---|---|---|---|---|
| $n_{det}(u)$ | 11 | 11 | 13 | 13 | 12 | 11 | 11 | 15 |

| $u$ | 8 | 9 | 10 | 11 | 12 | 13 | 14 | 15 |
|---|---|---|---|---|---|---|---|---|
| $n_{det}(u)$ | 11 | 11 | 8 | 7 | 11 | 14 | 8 | 8 |

The fault $f_0$ is detected by the set of input vectors $D(f_0) = \{9, 10, 11, 12, 13, 14, 15\}$. The smallest value of $n_{det}(u)$, where $u \in D(f_0)$, is obtained for $u = 11$, and it is equal to 7. This value implies that in the worst case, a test generated for $f_0$ will detect seven faults. The best case is if $u = 13$ is generated and detects 14 faults. However, we use the smallest value as the accidental detection index and set $ADI(f_0) = 7$.

The fault $f_2$ is detected by the set of input vectors $D(f_2) = \{4, 7, 13\}$. The smallest value of $n_{det}(u)$ is obtained for $u = 4$, and it is equal to 12. This value implies that in the worst case, a test generated for $f_2$ will detect 12 faults. We obtain $ADI(f_2) = 12$.

The fault $f_{15}$ is detected by the set of input vectors $D(f_{15}) = \{10, 14, 15\}$. All three input vectors have the same value of $n_{det}(u)$, equal to eight. Therefore, $ADI(f_{15}) = 8$.

## 3. Ordering the set of target faults

The accidental detection index is incorporated into the test generation process by ordering the set of target faults before test generation starts. We describe two pairs of fault orders based on the accidental detection index in this section. The first order in every pair is more suitable for the application of steepening the fault coverage curve. The second order in every pair is more suitable for dynamic test compaction. We also describe two orders that will be used later for comparison purposes.

From the definition of the accidental detection index, a test for a fault with a higher accidental detection index will detect more faults accidentally than a test for a fault with a lower accidental detection index. Consequently, it is expected that the test generation process will be improved if the faults are ordered by decreasing value of $ADI(f)$ such that faults with higher accidental detection indices appear earlier. We denote by $F_{decr} = <f_1, f_2, \cdots, f_n>$ the set of faults $F$ reordered such that $ADI(f_i) > ADI(f_j)$ for every $1 \le i < j \le n$.

We note that, in this work, the accidental detection indices are computed based on a set of input vectors $U$ that may not detect all the circuit faults. A fault $f$ that is not detected by $U$ is assigned $ADI(f) = 0$. We can treat the faults with $ADI(f) = 0$ in one of two ways. Since we cannot obtain a higher estimate of $ADI(f)$ from $U$, we can assume that the ability of a test for $f$ to accidentally detect other faults is low and keep $f$ at the end of the fault set. This is the approach taken in defining $F_{decr}$. A different approach is to place faults with $ADI(f) = 0$ at the beginning of the fault set since they are hard-to-detect, and they are not likely to be accidentally detected by tests for other faults. Consequently, it may be advantageous to target them first. To accommodate this view we define an ordered fault set, denoted by $F_{0decr}$, that includes the faults with zero accidental detection indices first, followed by the remaining faults in order of decreasing accidental detection index. The only difference between $F_{decr}$ and $F_{0decr}$ is that in $F_{decr}$ the faults with $ADI(f) = 0$ appear at the end, while in $F_{0decr}$ they appear at the beginning.

We point out that some of the faults with $ADI(f) = 0$ may be undetectable. For undetectable faults, their placement in the ordered fault set will not affect the test set size or the steepness of the fault coverage curve. Thus, we do not expect a difference between $F_{decr}$ and $F_{0decr}$ when the limited set of input vectors $U$ detects all the detectable circuit faults. When detectable faults have $ADI(f) = 0$, we expect $F_{decr}$ to yield a steeper fault coverage curve than $F_{0decr}$ since it follows more closely the accidental detection indices. We expect $F_{0decr}$ to yield a smaller test set since it first generates tests for the faults that are not likely to be accidentally detected by tests for other faults. Depending on the application, either $F_{decr}$ or $F_{0decr}$ should be used.

A dynamic ordering procedure will update the values of $n_{det}(u)$ and $ADI(f)$ as faults are included in the ordered fault set. This will imitate the detection of faults by the test generation procedure when it considers the ordered fault set. We denote the resulting ordered fault sets by $F_{dynm}$ and $F_{0dynm}$ (since values of $n_{det}(u)$ and $ADI(f)$ are updated dynamically during the ordering process). We demonstrate the construction of $F_{dynm}$ next by considering MCNC finite-state machine benchmark *lion*. The values of $n_{det}(u)$ for this circuit are shown in Table 1.

Initially, $F_{dynm} = \phi$. The highest accidental detection index is obtained for $f_{22}$ with $D(f_{22}) = \{7\}$ and $ADI(f_{22}) = 15$. We therefore include $f_{22}$ in $F_{dynm}$ first to obtain $F_{dynm} = <f_{22}>$. The value of $n_{det}(7)$ for input vector 7 can now be reduced by one, since $f_{22}$ does not need to be considered further. In general, we do not attempt to estimate which other faults will be detected by a test for a fault $f$, since this depends on the particular test generated for $f$. We only assume that $f$ will be dropped from the set of target faults after it is considered, and we update the values of $n_{det}(u)$ for every $u \in D(f)$. Following this we also update the value of $ADI(\hat{f})$ for every $\hat{f} \in F - F_{dynm}$.

The highest accidental detection index for *lion* is now obtained for $f_{18}$ with $D(f_{18}) = \{7, 13\}$ and $ADI(f_{18}) = 14$. We include $f_{18}$ in $F_{dynm}$ to obtain $F_{dynm} = <f_{22}, f_{18}>$. The values of $n_{det}(7)$ and $n_{det}(13)$ for input vectors 7 and 13 are reduced by one since $f_{18}$ does not need to be considered further.

The highest accidental detection index is now obtained for $f_{14}$ with $D(f_{14}) = \{2\}$ and $ADI(f_{14}) = 13$. We include $f_{14}$ in $F_{dynm}$ to obtain $F_{dynm} = <f_{22}, f_{18}, f_{14}>$. The value of $n_{det}(2)$ is reduced by one since $f_{14}$ does not need to be considered further.

The highest accidental detection index is now obtained for $f_{21}$ with $D(f_{21}) = \{7, 13\}$ and $ADI(f_{21}) = 13$. Here it is important to note that the original value of $ADI(f_{21})$ was 14, since $n_{det}(7)$ was originally 15 and $n_{det}(13)$ was originally 14. However, after $f_{22}$ was entered into $F_{dynm}$, $n_{det}(7)$ was reduced to 14, and after $f_{18}$ was entered into $F_{dynm}$ both values were reduced to 13, resulting in $ADI(f_{21}) = 13$. We include $f_{21}$ in $F_{dynm}$ to obtain $F_{dynm} = <f_{22}, f_{18}, f_{14}, f_{21}>$. The values of $n_{det}(7)$ and $n_{det}(13)$ are reduced by one.

Continuing in the same manner, we obtain an ordered fault set that takes into account the changes in $ADI(f)$ as faults are considered during test generation.

If the circuit has faults with $ADI(f) = 0$, we include them in $F_{dynm}$ at the end following the same process. We also define an ordered fault set $F_{0dynm}$, that includes the faults with $ADI(f) = 0$ first. We then follow the procedure demonstrated above to determine the order of the remaining faults in $F_{0dynm}$. Similar to the case of $F_{decr}$ and $F_{0decr}$, when detectable faults have $ADI(f) = 0$, we expect $F_{dynm}$ to yield a steeper fault coverage curve than $F_{0dynm}$ since it follows more closely the accidental detection indices. We expect $F_{0dynm}$ to yield a smaller test set since it first generates tests for the faults that are not likely to be accidentally detected.

For comparison purposes we also define the ordered fault set $F_{orig}$, where the faults are listed in their original order (given as part of the circuit description). In addition, we define the ordered fault set $F_{incr0}$ where the faults with $ADI(f) > 0$ are ordered by increasing accidental detection index, and the faults with $ADI(f) = 0$ are placed at the end. This fault order is expected to yield the worst results in terms of test set size. We confirm this point experimentally as another indication of the effectiveness of the accidental detection index.

## 4. Experimental results

In this section we present experimental results of test generation with different fault orders. The test generation procedure we use does not include any dynamic compaction heuristics. We denote an ordered fault set by $F_{ord}$. This can be $F_{orig}$, $F_{incr0}$, $F_{decr}$, $F_{0decr}$, $F_{dynm}$, or $F_{0dynm}$. We denote the test set computed for $F_{ord}$ by $T_{ord}$.

The circuits we consider are the combinational logic of ISCAS-89 and ITC-99 benchmarks. For ISCAS-89 benchmarks we consider irredundant versions of their combinational logic, referred to as *ircirc* where *circ* is the original circuit name.

The set of input vectors $U$ for the computation of the accidental detection indices consists of $N$ random input vectors, where $N$ is selected as follows. We initially include in $U$ 10,000 random input vectors. This is sufficient for achieving a high fault coverage for all the circuits considered. We simulate $U$ with fault dropping until all the vectors are simulated, or until approximately 90% of the circuit faults are detected. If this happens after $N$ vectors are simulated, we keep in $U$ only the first $N$ vectors. This ensures that accidental detection indices are computed for a sufficiently large percentage of the faults. At the same time, it typically does not take a large number of input vectors to reach 90% fault coverage in the circuits considered. Consequently, the accidental detection index can be computed efficiently. The process can be speeded up further by removing from consideration vectors in $U$ that do not detect any new faults during fault simulation with fault dropping of $U$.

The accidental detection indices are shown in Table 4. The results relevant to dynamic test compaction are reported in Table 5. Run times are reported in Table 6. In Figure 1 and Table 7 we provide results for the application of steepening the fault coverage curve.

**Table 4: Accidental detection index**

| circuit | inp | vec | ADI min | ADI max | ratio |
|---|---|---|---|---|---|
| irs208 | 19 | 370 | 11 | 31 | 2.82 |
| irs298 | 17 | 109 | 19 | 61 | 3.21 |
| irs344 | 24 | 37 | 42 | 71 | 1.69 |
| irs382 | 24 | 82 | 10 | 76 | 7.60 |
| irs400 | 24 | 74 | 24 | 101 | 4.21 |
| irs420 | 35 | 3523 | 16 | 56 | 3.50 |
| irs510 | 25 | 87 | 31 | 66 | 2.13 |
| irs526 | 24 | 190 | 27 | 102 | 3.78 |
| irs641 | 54 | 192 | 50 | 85 | 1.70 |
| irs820 | 23 | 1328 | 4 | 41 | 10.25 |
| irs953 | 45 | 2030 | 60 | 136 | 2.27 |
| irs1196 | 32 | 1211 | 36 | 117 | 3.25 |
| irs5378 | 214 | 358 | 753 | 973 | 1.29 |
| irs13207 | 699 | 3110 | 2113 | 2668 | 1.26 |

In Table 4, after the circuit name we show the number of inputs, and the number of random input vectors included in $U$. Under column *ADI* we show the value of $ADI_{min} = \min\{ADI(f): f \in F_U\}$, the value of $ADI_{max} = \max\{ADI(f): f \in F_U\}$, and the ratio $ADI_{max}/ADI_{min}$. For the faults left undetected by $U$, the accidental detection index is zero by definition. The minimum and maximum accidental detection indices are computed considering only detected faults.

It can be seen from Table 4 that the differences between the smallest and the largest accidental detection indices are significant. Therefore, the accidental detection index is expected to make a difference in the test set size and in the steepness of the fault coverage curve if used as a fault ordering heuristic.

**Table 5: Test generation**

| | tests | | | |
|---|---|---|---|---|
| circuit | orig | dynm | 0dynm | incr0 |
| irs208 | 42 | 33 | 34 | 41 |
| irs298 | 43 | 36 | 37 | 43 |
| irs344 | 28 | 26 | 25 | 31 |
| irs382 | 42 | 45 | 35 | 47 |
| irs400 | 37 | 37 | 35 | 44 |
| irs420 | 70 | 62 | 60 | 71 |
| irs510 | 65 | 67 | 66 | 74 |
| irs526 | 75 | 74 | 64 | 86 |
| irs641 | 74 | 66 | 59 | 77 |
| irs820 | 149 | 130 | 125 | 179 |
| irs953 | 110 | 106 | 99 | 120 |
| irs1196 | 179 | 162 | 153 | 209 |
| irs5378 | 254 | 240 | 224 | - |
| irs13207 | 411 | 397 | 342 | - |
| average | 112.8 | 105.8 | 97.0 | - |

In Table 5 we show the results of test generation using $F_{orig}$, $F_{dynm}$, $F_{0dynm}$ and $F_{incr0}$. We consider $F_{orig}$ and $F_{incr0}$ (for some of the circuits) for comparison purposes. We do not consider $F_{decr}$ and $F_{0decr}$ since $F_{dynm}$ and $F_{0dynm}$ proved to be better. We show the test set sizes for the different ordered fault sets under columns *orig*, *dynm*, *0dynm* and *incr0*, respectively.

From Table 5 it can be seen that both $F_{dynm}$ and $F_{0dynm}$ reduce the test set size compared to using $F_{orig}$. Using $F_{incr0}$ increases the test set size. This supports the effectiveness of the accidental detection index and the ordering of faults by decreasing values of the index. Using $F_{0dynm}$ results in the smallest test sets overall. This is due to the fact that hard-to-detect faults (faults with $ADI(f) = 0$) are placed first in $F_{0dynm}$. We expect $F_{dynm}$ to provide a steeper fault coverage curve if the faults with $ADI(f) = 0$ do not have high accidental detection indices. The results reported later address this point.

Next, we report the effect of fault ordering on the run times of test generation as follows. Let the test generation time for an ordered fault set $F_{ord}$ be $RT_{ord}$. We report the relative run times $RT_{ord}/RT_{orig}$ for $F_{orig}$, $F_{dynm}$ and $F_{0dynm}$ in Table 6.

From Table 6 it can be seen that the test generation time for $F_{dynm}$ is between 0.45 and 2.21 times the run time for $F_{orig}$, while for $F_{0dynm}$ it is between 0.53 and 1.96 times the run time for $F_{orig}$. The average run times are 1.14 and 0.98 of the run time for $F_{orig}$, respectively. In many cases the run time for $F_{dynm}$ and $F_{0dynm}$ is reduced compared to that obtained for $F_{orig}$. This is an advantage compared to other dynamic compaction heuristics [1]-[4] that typically increase the run time significantly.

**Table 6: Relative run times**

| | t.gen | | |
|---|---|---|---|
| circuit | orig | dynm | 0dynm |
| irs208 | 1.00 | 1.18 | 0.76 |
| irs420 | 1.00 | 1.88 | 1.10 |
| irs510 | 1.00 | 1.06 | 0.77 |
| irs641 | 1.00 | 0.96 | 1.27 |
| irs820 | 1.00 | 0.68 | 0.73 |
| irs953 | 1.00 | 1.03 | 0.93 |
| irs1196 | 1.00 | 2.21 | 1.96 |
| irs5378 | 1.00 | 0.45 | 0.53 |
| irs13207 | 1.00 | 0.83 | 0.78 |
| average | 1.00 | 1.14 | 0.98 |

To demonstrate the effect of fault ordering on the steepness of the fault coverage curve, we plot the increase in fault coverage as the number of test vectors increases during test generation for *irs*420. We use the three ordered fault sets, $F_{orig}$ (for which the data points are given by *o*'s), $F_{dynm}$ (for which the data points are given by *d*'s) and $F_{0dynm}$ (for which the data points are given by *z*'s). The curve obtained for *irs*420 is shown in Figure 1. The x-axis is the number of tests as a percentage of the largest test set size. The y-axis is the fault coverage.

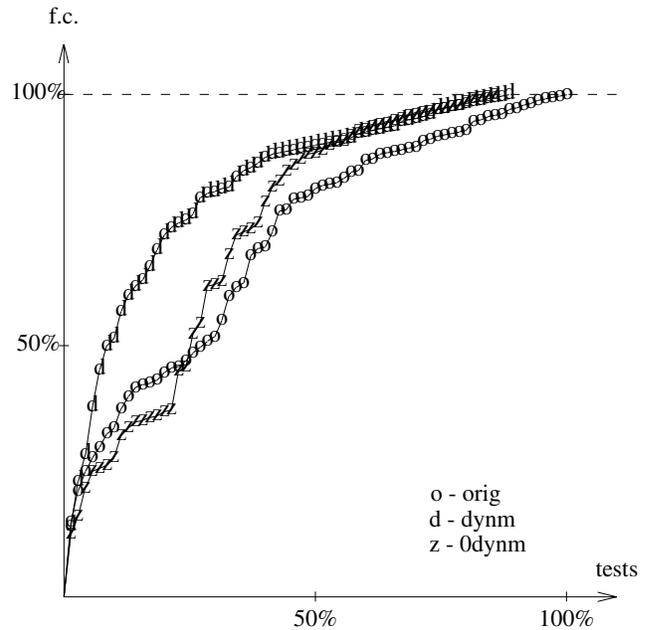

**Figure 1: Fault coverage curve for irs420**

In order to provide results for additional circuits in a concise way we compute the expected value of the number of tests that need to be applied in order to detect a faulty chip. This parameter is related to the steepness of the fault coverage curve as follows. With a steeper curve, the average number of tests it takes to detect a fault is lower. As a result, faults (and defects) are detected earlier during the test application process.

Let $T_{ord} = <t_{ord1}, t_{ord2}, \cdots, t_{ordk_{ord}}>$. Let the number of faults detected by $<t_{ord1}, t_{ord2}, \cdots, t_{ordi}>$ be $n_{ord}(i)$ (the values of $n_{ord}(i)$ are used for plotting the fault coverage curve in Figure 1). We have $n_{ord}(1)$ faults detected after the application of a single test, $n_{ord}(2) - n_{ord}(1)$ faults detected after the application of two tests, and so on. By defining $n_{ord}(0) = 0$, we have that $n_{ord}(i) - n_{ord}(i-1)$ faults are detected after the application of $i$ tests, for $1 \leq i \leq k$. The average number of tests that need to be applied for a fault to be detected is defined as

$$AVE_{ord} = \frac{\sum_{i=1}^{k_{ord}} i \cdot [n_{ord}(i) - n_{ord}(i-1)]}{n_{ord}(k_{ord})}.$$

In Table 7 we report the values of $AVE_{ord}$ normalized to $AVE_{orig}$, i.e., we show the values of $AVE_{ord}/AVE_{orig}$ for $F_{orig}$, $F_{dynm}$ and $F_{0dynm}$. A lower value in Table 7 implies a steeper fault coverage curve, where a fault is expected to be detected earlier during the test application process.

**Table 7: Steepness of fault coverage curves**

| circuit | AVEord orig | dynm | 0dynm |
|---|---|---|---|
| irs208 | 1.000 | 0.644 | 0.833 |
| irs298 | 1.000 | 0.796 | 0.870 |
| irs344 | 1.000 | 1.023 | 0.853 |
| irs382 | 1.000 | 1.081 | 0.918 |
| irs400 | 1.000 | 0.963 | 0.973 |
| irs420 | 1.000 | 0.614 | 0.894 |
| irs510 | 1.000 | 0.834 | 0.895 |
| irs526 | 1.000 | 0.964 | 0.957 |
| irs641 | 1.000 | 0.830 | 0.807 |
| irs820 | 1.000 | 0.825 | 0.875 |
| irs953 | 1.000 | 0.998 | 1.191 |
| irs1196 | 1.000 | 0.790 | 0.868 |
| irs5378 | 1.000 | 0.865 | 0.873 |
| irs13207 | 1.000 | 0.960 | 0.767 |
| average | 1.000 | 0.870 | 0.898 |

From Figure 1 and Table 7 it can be seen that in most of the cases considered, the fault coverage curve obtained for $F_{dynm}$ is steeper than the other two even though the size of the test set obtained for $F_{0dynm}$ is smaller. The curve for $irs420$ demonstrates a case where placing the faults with $ADI(f) = 0$ (the faults that are undetected by $U$) at the beginning of $F_{0dynm}$ causes fewer accidental detections to be obtained for the first few test vectors than when these faults are placed at the end as in $F_{dynm}$, or when they are placed arbitrarily as in $F_{orig}$. The last row of Table 7 shows that, on the average, the expected value of the number of tests applied before a faulty chip is detected is reduced by 13% when the ordered fault set $F_{dynm}$ is used.

## 5. Concluding remarks

We defined the accidental detection index $ADI(f)$ of a fault $f$ as the minimum number of faults that will be detected by a test generated for $f$. We provided a method for computing an estimate of the accidental detection index based on the results of simulating a set of input vectors $U$. We used the accidental detection index to order the set of target faults before starting test generation. Fault ordering was done such that a fault with a higher accidental detection index appeared earlier in the ordered fault set. The ordered fault set was shown to be effective for dynamic test compaction, and for obtaining a test set with a steep fault coverage curve. For the application of dynamic test compaction it was shown that it is better to place faults with $ADI(f) = 0$ at the beginning of the ordered set of faults, and order the remaining faults by decreasing accidental detection index. For the application of steepening the fault coverage curve it was shown that it is better to order all the faults by decreasing accidental detection index, placing faults with $ADI(f) = 0$ at the end of the ordered set.